\documentclass[12pt]{article}
\usepackage[colorlinks,linkcolor=blue,citecolor=blue,urlcolor=blue,bookmarks,bookmarksnumbered]{hyperref}
\usepackage{amsmath,amssymb,amsfonts}
\usepackage[paper=letterpaper,margin=1.0in]{geometry}
\usepackage{graphicx,xcolor}
\usepackage{cite}
\usepackage{color}

\begin{document}
\thispagestyle{empty}

\begin{center}

\vspace{2.0cm}
{\large \bf Dark Matter, Dark Energy, and Fundamental Acceleration}
\vspace{1.0cm}

\renewcommand{\thefootnote}{\fnsymbol{footnote}}

{\bf
Douglas Edmonds${}^{1}$\footnote{\tt bde12@psu.edu, Corresponding author},
Djordje Minic${}^{2}$\footnote{\tt dminic@vt.edu},
and
Tatsu Takeuchi${}^{2}$\footnote{\tt takeuchi@vt.edu}
}

\vskip 0.5cm

{\footnotesize\it
${}^1$Department of Physics, Penn State Hazleton, Hazleton, PA 18202, U.S.A. \\
${}^2$Department  of Physics, Virginia Tech, Blacksburg, VA 24061, U.S.A. \\
${}$ \\
}

\vskip 1.0cm

\begin{abstract}\unskip\noindent
\begin{quotation}\noindent
We discuss the existence of an acceleration scale in galaxies and galaxy clusters and its relevance for the nature of dark matter. The presence of the same acceleration scale found at very different length scales, and in very different astrophysical objects, strongly supports the existence of a \emph{fundamental} acceleration scale governing the observed gravitational physics. We comment on the implications of such a fundamental acceleration scale for constraining cold dark matter models as well as its relevance for structure formation to be explored in future numerical simulations.
 
\end{quotation}
\end{abstract}

\vspace{1cm}







\end{center}

\noindent This essay received an honorable mention in the Gravity Research Foundation 2020
Awards for Essays on Gravitation competition.

\renewcommand{\thefootnote}{\arabic{footnote}}

\newpage
\setcounter{page}{1}
\pagestyle{plain}

The mystery of the nature of dark matter is closely intertwined with the problem
of structure formation in the Universe, namely how structures
such as globular clusters ($\sim\!10$ pc), 
galaxies ($\sim\!10$ kpc), and galaxy clusters ($\sim\!10$ Mpc)
evolved, or are currently evolving, at various different length-scales.

According to the $\Lambda$CDM model of cosmology, 
our Universe started out with almost uniform distributions of cold dark matter (CDM) and baryonic matter.
The properties of CDM are completely unknown, and the only assumptions are that it is non-relativistic (Cold) 
and that it only interacts with the baryonic matter (and itself) gravitationally.
Small fluctuations in the initial dark matter distribution evolve into filaments and voids, as seen in the famous ``Millennium Simulation,''
forming gravitational potential wells into which the baryonic matter collapse, after the ``last-scattering'' of photons,
collecting into clumps of gas which eventually evolve into the structures we see today \cite{Peacock,Weinberg}.

Our understanding of this structure formation process is incomplete.
Observations such as the Hubble Deep Field have allowed us to peak back in time
to the early stages of structure formation, with the surprising revelation that galaxies started to form
earlier than previously envisioned. 
However, the observations of faint and/or far-away objects are limited by the sensitivity and resolution of the detectors
available, and we have yet to see the full history of structure formation directly \cite{Longair,Combes}.

$N$-body simulations also provide us with insights into structure formation.
Yet, the currently available computer power only allows simulations of very coarsely grained 
and roughly modeled systems, 
especially when simulating baryonic matter, due to the vastly different time- and length-scales 
of the processes involved.
Nevertheless, there are hints that $\Lambda$CDM may be insufficient to fit all observations
toward the galactic scale and below \cite{Navarro:1995iw,Governato:2006cq},
indicating that we may need to go beyond the $\Lambda$CDM paradigm \cite{Bull:2015stt}.


A curious observational fact which is not often emphasized (or only emphasized in the limited context of MOND \cite{Milgrom:1983ca}), 
is that the dark- and baryonic-matter distributions in galaxies today (redshift $z \sim 0$)
are strongly correlated via an acceleration scale which appears to be universal to all spiral galaxies.
A correlation between the total dark matter and total baryonic matter in a region surrounding a galaxy is to be
expected in the $\Lambda$CDM scenario, but given that galaxies are thought to have gone through various phases
including starbursts, emission of gasses, and multiple mergers during their evolutionary histories \cite{Longair,Combes}, 
a dynamic correlation involving a universal acceleration scale is surprising.

The existence of this acceleration scale universal to spiral galaxies 
is implicit in the
Tully-Fisher relation (TFR, 1977 \cite{Tully:1977fu}). 
Originally noted as an empirical relation between the absolute optical luminosity $L$ of spiral galaxies and the
width of their \textsc{Hi} 21~cm lines, 
it has since been reformulated into the \textit{baryonic} Tully-Fisher
relation (BTFR) which posits a relation between the total baryonic mass of the galaxy $M_{bar}$
(the sum of masses of the stars and gas)
and the asymptotic rotational velocity $v_\infty$ of the form \cite{Freeman:1999,McGaugh:2000sr}
\begin{equation}
M_{bar} \,\propto\, v_\infty^4 \;.
\label{BTFR}
\end{equation}
This relation has been studied intensively by various groups (see Table~3 in \cite{TorresFlores:2011uc})
and has been found to apply with smaller scatter than the original TFR 
to both HSB (high surface brightness) and LSB (low surface brightness) spiral galaxies, 
as well as irregular \cite{TorresFlores:2011uc} and dwarf irregular galaxies \cite{Iorio_2017},
covering $\sim 5$ decades of mass scale: $M_{bar} = 10^{6\sim 11}M_\odot$. (See Fig.~23 of \cite{Iorio_2017}).
%
The slope of Eq.~\eqref{BTFR} is the universal acceleration scale
\begin{equation}
a_{\S} \,=\, \frac{v_\infty^4}{G M_{bar}}\;,
\end{equation}
where $G$ is Newton's gravitational constant.
Scalewise, galactic rotation curve data yields $a_\S \approx 10^{-10}\,\mathrm{m/s^2}$ \cite{McGaugh:2011ac}.
Note that $v_\infty$ is a proxy for the centripetal acceleration of the object due to
the total mass of the galaxy, including contributions from both baryonic and dark matter.
Thus, the BTFR implies a universal correlation between the observed acceleration $a_{obs}$ (as deduced from $v_\infty$)
and the expected acceleration $a_{bar}$ from the baryonic matter $M_{bar}$ alone toward the outskirts of
spiral galaxies, and this correlation seems to be governed by the universal acceleration scale $a_\S$.
We would like to emphasize here we are quoting an observational fact, which is independent of
any theoretical assumptions or biases one may have about the dynamics responsible.\footnote{%
In particular, we are NOT pushing MOND.}

For HSB spirals, universal correlations between the $a_{obs}$ and $a_{bar}$ have been noted by various authors at intermediate radii as well,
though not necessarily expressed as a relation between the two \cite{Rubin1985ApJ289,Persic:1991}.
McGaugh et al. derive such a relation in Ref.~\cite{McGaugh:2016leg}
using 2693 data points at various radii from 153 late type galaxies (LTGs = spirals and irregulars) 
in the SPARC database \cite{Lelli:2016}.
These data points cover a range of $a_{bar}$ from about $10^{-8}\,\mathrm{m/s^2}$ down to about $10^{-12}\,\mathrm{m/s^2}$ 
and the relation between $a_{obs}$ and $a_{bar}$ is investigated. 
It is shown that the data fall nicely on the so-called Radial Acceleration Relation (RAR) curve
{(or, alternatively, the Mass-Discrepancy Acceleration Relation (MDAR) \cite{McGaugh:2004})}:
\begin{equation}
a_{obs} 
\;=\; \frac{a_{bar}}{1-e^{-\sqrt{a_{bar}/a_{\S}}}}
\;.
\label{MDAR}
\end{equation}
The best-fit value of the acceleration parameter $a_{\S}$ in this expression was found to be $a_{\S} = (1.2 \pm 0.02(\text{stat})\pm 0.24(\text{syst})) \times 10^{-10}\,\mathrm{m/s^2}$. 
It is straightforward to show that MDAR implies BTFR.
Salucci in Ref.~\cite{Salucci:2016vxb} confirms the MDAR by translating the results from
Persic, Salucci et al. in Ref.~\cite{Persic:1991},
which used roughly ten times more data than \cite{McGaugh:2016leg}
from 967 galaxies, and also different analysis techniques.
(See also Ref.~\cite{Salucci:2018eie}.\footnote{%
Though McGaugh et al. \cite{McGaugh:2016leg} and Salucci \cite{Salucci:2016vxb}
agree completely on what the data say, they disagree on the implications for the properties of dark matter.
})
Di Paolo, Salucci, and Fontaine in \cite{DiPaolo:2018mae} further 
extend the analysis to LSB galaxies and dwarf-spirals and find that 
while data points toward the center of these galaxies do not fall on the MDAR,
those toward the outskirts do, as expected from the BTFR.
Thus MDAR may be an indication that $a_\S$ plays an extra role in the dynamics of HSB galaxies.

The relation for elliptical galaxies and other pressure supported systems that parallels the TFR for spirals is 
the Faber-Jackson relation (FJR, 1976 \cite{Faber:1976sn}).
Again, the FJR was originally noted as a proportionality between the absolute optical luminosity $L$ of 
elliptical galaxies and a power of the line-of-sight (los) velocity dispersion $\sigma$ \cite{Faber:1976sn}. 
The corresponding \textit{baryonic} Faber-Jackson relation (BFJR) is
\begin{equation}
M_{bar} \;\propto\; \sigma^n\;.
\end{equation}
Though it was originally proposed that $L\propto\sigma^4$, subsequent analyses of elliptical galaxy data have found the
power $n$ to be anywhere from about 3 to 5, depending on the dataset and analysis method used \cite{Kormendy:1982}, and as low as 2 for dwarf galaxies \cite{Cody:2009}.
The BFJR can also be fit to globular cluster data, but there, there is considerable scatter 
so the meaning of the fit by itself is unclear \cite{Nella-Courtois:1999}.
However, as was pointed out by Farouki, Shapiro, and Duncan in 1983 \cite{Farouki:1983}, 
when data for elliptical galaxies \cite{Tonry:1981} 
and globular clusters \cite{Peterson:1975,Illingworth:1976} are plotted together,
the data is seen to cluster along an $n=4$ line spanning 8 decades of mass scale:
$M_{bar} = 10^{4\sim 12}M_\odot$. 
And just as in the BTFR, $n=4$ would immediately imply the existence of an acceleration scale
\begin{equation}
a_{\varnothing} \;=\; \dfrac{\sigma^4}{G M_{bar}}\;,
\label{eqn:a_elliptical}
\end{equation}
which is universal among pressure supported systems including elliptical galaxies and globular clusters.
Furthermore, the slope found in \cite{Farouki:1983} from a fit to \cite{Tonry:1981}
implies $a_{\varnothing} \sim O(10^{-10}\,\mathrm{m/s^2})$.
%
%
Thus, the BTFR and the BFJR together seem to point to the existence of an acceleration scale
of order 
$10^{-10}\,\mathrm{m/s^2} \equiv a_0$, which is universal to both rotation and pressure supported systems
including and below the galaxy scale.\footnote{There are recent indications that the so-called "ultra-diffuse galaxies" (UDGs) may deviate significantly from observed correlations such as BTFR and BFJR. Dark matter content seems to extend from too much \cite{Forbes:2020} to too little \cite{vanDokkum:2019, Pina:2019} in these objects. While our understanding of UDGs is not yet clear, they could represent cases not included in the ``universality class'' that we are proposing. }
Again, we emphasize that we are discussing observational facts, independent of
any theoretical considerations.

Where could this universal acceleration scale $a_0$ be coming from?
If we are to keep the theory of gravity intact (no MOND!), two possibilities come to mind:
\begin{enumerate}
\vspace{-0.3cm}
\item The scale is emergent in the $\Lambda$CDM model.  
It is an indication of the robustness of the galaxy formation process and will consistently emerge
in numerical simulations once the evolutions of both dark- and baryonic-matter are properly included. 
This point of view is discussed in Ref.~\cite{Kaplinghat:2001me}.
See also Ref.~\cite{Navarro:2016bfs} and \cite{Glowacki_2019}.

\vspace{-0.2cm}
\item The existence of this universal acceleration scale hints at a new heretofore unknown property of CDM.
Noting that the value of the acceleration scale can be related to the Hubble scale $H_0$ via
$a_0 \approx cH_0/(2\pi)$, and noting further that $\Lambda\sim 3H_0^2$,
could it be pointing to a correlation between dark matter and dark energy? 
See, for instance, Refs.~\cite{Khoury:2014tka,Verlinde:2017}.
\end{enumerate}
\vspace{-0.3cm}
We have also considered the second possibility in previous works \cite{Edmonds:2013hba}, where
we confront the data with the modified dark matter (MDM) proposal of Ref.~\cite{Ho:2010ca} by Ho, Ng, and one of us.
In MDM, the CDM quanta is informed of the cosmological constant $\Lambda$ via gravitational thermodynamics.

At this point in time, Occam's razor would favor the first option, given that 
recent $N$-body simulations of galaxy evolution do seem to yield results 
consistent with MDAR \cite{Navarro:2016bfs} and BTFR \cite{Glowacki_2019}.
However, what if the same acceleration scale appeared in structures at vastly larger length-scales as well?

Let us look at galaxy-clusters.  
In Ref.~\cite{Zhang:2010qk}, Zhang et al. analyze 62 
of the 64 galaxy clusters in the HIFLUGCS database \cite{HIFLUGCS}, 
and for each cluster provide the los velocity dispersion $\sigma$ of the member galaxies (13439 galaxies total),
and the total baryonic mass $M_{bar}$ inside the cluster radius $r_{500}$
($r_{500}$ is the radius within which the baryonic mass density is 500 times the critical density), 
which was estimated from the gas mass $M_{gas}$ obtained from X-rays observations.
Note that $\sigma$ and $M_{bar}$ are determined from independent observations. 
Using this data, we plot $\sigma^4/(Ga_0)$ against $M_{bar}$ in Figure~1.
We see that the 62 data points (black circles) cluster around $\sigma^4/(Ga_0) = M_{bar}$ (black solid line),
indicating that $a_0$ also appears in the dynamics of galaxy clusters.


\begin{figure}[t]
\begin{center}
\includegraphics[width=10cm]{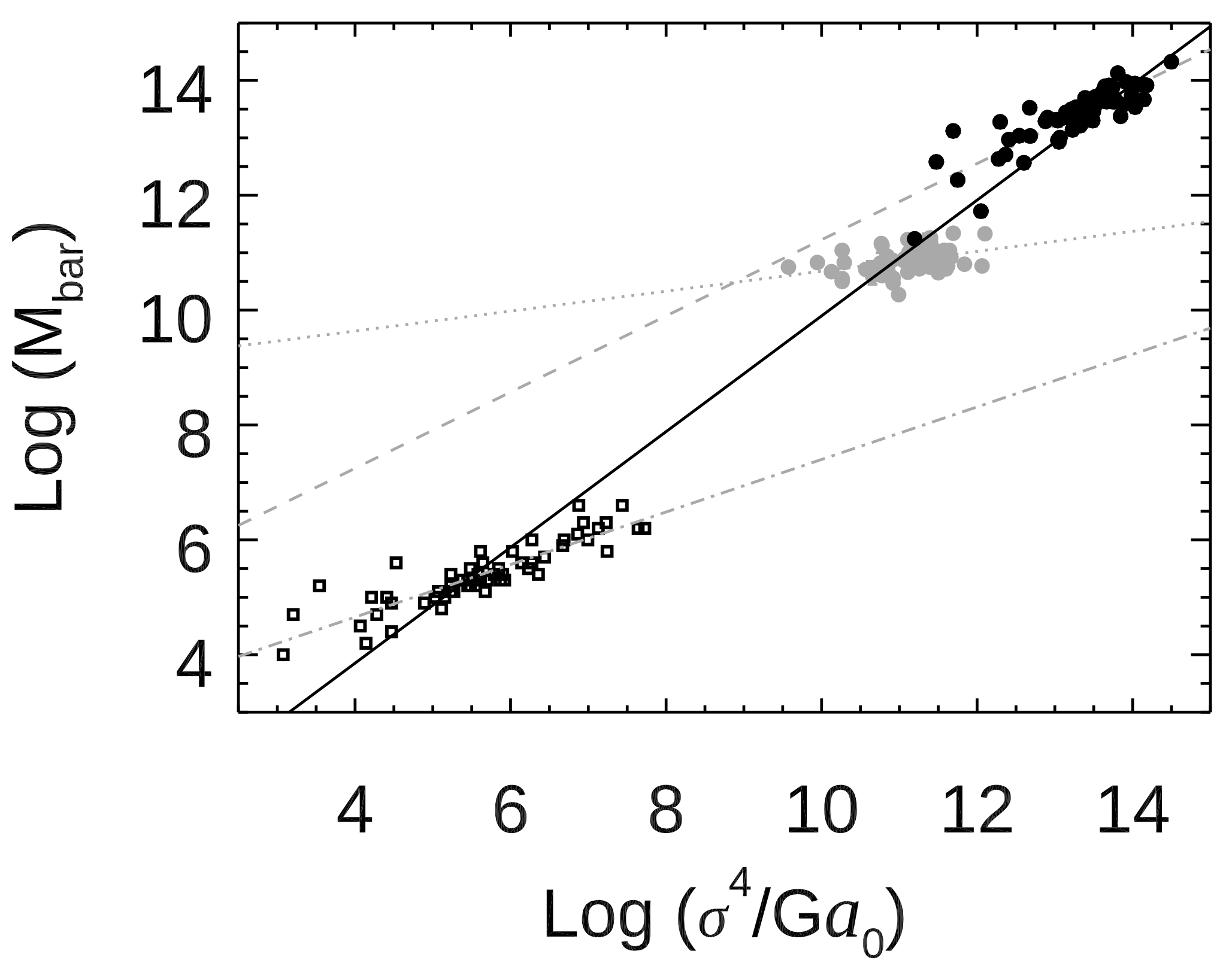}
\caption{The Faber-Jackson relation including galaxy clusters, elliptical galaxies, and globular clusters. The black and grey circles correspond, respectively, to clusters and elliptical galaxies. The open squares represent data from globular clusters. The dashed line is the best fit to the cluster data, the dotted and dot-dashed lines are the best-fit to the elliptical galaxies alone and the globular clusters alone, respectively. The solid line is the best-fit to all of the data taken together.}
\label{fig:globulars}
\end{center}
\end{figure}


Figure~1 also shows 56 data points for elliptical galaxies (gray circles),
and 56 data points for globular clusters (open squares),
following the analysis of \cite{Farouki:1983}.
Velocity dispersions and mass estimates for the elliptical galaxies were taken from Ref.~\cite{Belli:2014}, and those for the globular clusters were taken from Ref.~\cite{Meylan:1993yd}. 
While the mass estimates for globular clusters are not entirely independent of velocity dispersion measurements, 
they were based upon the King model for globular clusters (see \cite{King:1968} and references therein) rather than relying on observed correlations for which dispersion is just one of the model parameters. 
With galaxy clusters, elliptical galaxies, and globular clusters all included in the plot, we see that
the BFJR with $n=4$ applies to 10 decades of mass scale: $M_{bar} = 10^{4\sim 14}M_\odot$.

We note that the existence of the BFJR and $a_0$ in galaxy cluster data had been pointed out as early as 1994 by Sanders 
in Ref.~\cite{Sanders:1994}, in which the MONDian virial theorem, which includes $a_0$, 
was used to relate $\sigma$ and $M_{bar}$.
The analysis was extended to include globular clusters in Ref.~\cite{Sanders:2010aa}
to posit a ``universal Faber-Jackson relation'' similar to Figure 1.
See also Refs.~\cite{Duranzo2017,Bilek2019}.
Our current analysis indicates that the appearance of the BFJR straddling vastly different scales
does not depend on any dynamics, Newtonian or otherwise.
It is purely an observational fact.

What are we to make of this result?
The evolution of galaxy clusters again is thought to involve various processes
through which it is difficult to imagine that any ``memory'' of $a_0$ of each galaxy in the cluster
would be maintained or transferred to the cluster intact.
If the presence of $a_0$ is due to dark matter, what type of quanta must it be?

Let us speculate: 
The BTFR and BFJR imply that the dark matter mass profile knows about the baryonic mass profile, 
the acceleration scale $a_0$, 
and the inertial properties of tracers in the gravitational potential. 
The dark-matter quanta are therefore inherently \textit{non-local}. 
A possible candidate for such quanta are \textit{metaparticles}, 
which are found in the long-distance limit of an intrinsically non-commutative formulation of quantum gravity/string theory~\cite{Freidel:2013zga}.
In this context, dark matter is correlated to the visible baryonic matter via a fundamental length scale that can
be related to dark energy (or the cosmological constant, $\Lambda \sim 3{H_0}^2$) \cite{rBHM8, Minic:2020oho}.
Thus, the BTFR and BFJR could be hints that the CDM quanta is something quite exotic.

On the other hand, it could still be that $a_0$ will emerge in galaxy-clusters within the
standard $\Lambda$CDM framework, just as it seems to do for galaxies \cite{Navarro:2016bfs,Glowacki_2019}.
Forthcoming analysis of Cluster-EAGLE simulations~\cite{Barnes:2017} may help us better understand the extent to which 
$\Lambda$CDM can accommodate the existence of $a_0$ in both galaxies and galaxy clusters. 

Colliding clusters may also offer further clues. 
We have thus far considered systems that are mostly or completely relaxed. 
If $a_0$ emerges in these relaxed systems as the end product of the interaction of baryonic matter and dark matter, 
we would expect that there will be no evidence for it in colliding clusters. 
However, the presence of $a_0$ in colliding clusters would be very strong evidence that $a_0$ is fundamental to nature.
In this sense, our discussion naturally relates to
the observationally supported proposal for dark matter quanta that are sensitive to dark energy~\cite{Ho:2010ca}. 
This, in turn, should have important implications for structure formation and it should be explored in numerical simulations.

\noindent
{\bf Acknowledgments:} 
We thank P. Berglund, L. Freidel, S. Horiuchi, T. Hubsch, J. Kowalski-Glikman and R. Leigh 
for discussions.
DM and TT are supported in part by the US Department of Energy (DE-SC0020262).
DM is also supported in part by the Julian Schwinger Foundation, 
and TT by the US National Science Foundation (NSF Grant 1413031).

\footnotesize\baselineskip=2.5ex
%

\begin{thebibliography}{100}


\bibitem{Peacock}
J. A. Peacock,
\textit{``Cosmological Physics,''}
Cambridge University Press, 1998.

\bibitem{Weinberg}
S. Weinberg,
\textit{``Cosmology,''}
Oxford University Press, 2008.



\bibitem{Longair}
M. S. Longair,
\textit{``Galaxy Formation,''}
2nd edition, Springer 2008.

\bibitem{Combes}
F. Combes,
\textit{``Mysteries of Galaxy Formation,''}
Springer Praxis Books 2010.


\bibitem{Navarro:1995iw} 
  J.~F.~Navarro, C.~S.~Frenk and S.~D.~M.~White,
  Astrophys.\ J.\  {\bf 462}, 563 (1996)


\bibitem{Governato:2006cq} 
  F.~Governato et al.,
  Mon.\ Not.\ Roy.\ Astron.\ Soc.\ \textbf{374}, 1479 (2007),
  Mon.\ Not.\ Roy.\ Astron.\ Soc.\ \textbf{422}, 1231 (2012).

\bibitem{Bull:2015stt} 
  P.~Bull {\it et al.},
  Phys.\ Dark Univ.\  \textbf{12}, 56 (2016).

\bibitem{Milgrom:1983ca}
M. Milgrom, Astrophys. J. \textbf{270}, 365 (1983); 371 (1983); 384 (1983);
\textbf{306}, 9 (1986).





\bibitem{Tully:1977fu}
R. B. Tully and J. R. Fisher, Astron. Astrophys. \textbf{54}, 661
(1977).

\bibitem{Freeman:1999} 
K. C. Freeman, in {\it The Low Surface Brightness Universe},
Astronomical Society of the Pacific Conference Series,
Vol. 170, edited by J. I. Davies, C. Impey, and S. Phillipps
(San Francisco, California, 1999) pp. 3-8.

\bibitem{McGaugh:2000sr}
S. S. McGaugh, J. M. Schombert, G. D. Bothun, and
W. J. G. de Blok, Astrophys. J. \textbf{533}, L99 (2000).

\bibitem{TorresFlores:2011uc}
S. Torres-Flores, B. Epinat, P. Amram, H. Plana, and C. Mendes de Oliveira,
Mon. Not. Roy. Astron. Soc. \textbf{416}, 1936 (2011).

\bibitem{Iorio_2017}
G. Iorio, F. Fraternali, C. Nipoti, E. Di Teodoro, J. I. Read, and G. Battaglia,
Mon. Not. Roy. Astron. Soc. \textbf{466}, 4159 (2017).

\bibitem{McGaugh:2011ac}
S. McGaugh, Astron. J. \textbf{143}, 40 (2012).

\bibitem{Rubin1985ApJ289} 
V. C. Rubin, D. Burstein, J. Ford, W. K., and N. Thonnard,
Astrophys. J. \textbf{289}, 81 (1985).

\bibitem{Persic:1991}
M. Persic and P. Salucci, Astrophys. J. \textbf{368}, 60 (1991),
M. Persic, P. Salucci, and F. Stel, 
Mon. Not. Roy. Astron. Soc. \textbf{281}, 27 (1996),
P. Salucci et al, 
Mon. Not. Roy. Astron. Soc. \textbf{378}, 41 (2007).


\bibitem{McGaugh:2016leg}
S. McGaugh, F. Lelli, and J. Schombert, 
Phys. Rev. Lett. \textbf{117}, 201101 (2016).

\bibitem{Lelli:2016}
F. Lelli, S. S. McGaugh, and J. M. Schombert, 
Astron. J. \textbf{152}, 157 (2016).

\bibitem{McGaugh:2004}
S. McGaugh, Astophys. J. {\bf609}, 652 (2004)

\bibitem{Salucci:2016vxb} 
P. Salucci, (2016), arXiv:1612.08857 [astro-ph.GA].

\bibitem{Salucci:2018eie}
P. Salucci, 
Found. Phys. \textbf{48}, 1517 (2018);
Astron. Astrophys. Rev. \textbf{27}, 2 (2019).

\bibitem{DiPaolo:2018mae}
C. Di Paolo, P. Salucci, and J. P. Fontaine,
Astrophys. J. \textbf{873}, 106 (2019).


\bibitem{Faber:1976sn}
S. M. Faber and R. E. Jackson, Astrophys. J. \textbf{204}, 668 (1976).


\bibitem{Kormendy:1982}
J. Kormendy, in {\it Morphology and Dynamics of Galaxies},
Saas-Fee Advanced Course, Vol. 12, edited by L. Martinet
and M. Mayor (Observatoire de Geneve, Sauverny,
Switzerland, 1982) pp. 113-288.

\bibitem{Cody:2009}
A. M. Cody, D. Carter, T. J. Bridges, B. Mobasher, B. M. Poggianti,
Mon. Not. Roy. Astron. Soc. {\bf 396}, 1647 (2009)

\bibitem{Nella-Courtois:1999}
H. di Nella-Courtois, P. Lanoix, G. Paturel,
Mon. Not. Roy. Astron. Soc. {\bf 302}, 587 (1999)

\bibitem{Farouki:1983}
R. T. Farouki, S. L. Shapiro, and M. J. Duncan,
Astrophys. J. \textbf{265}, 597 (1983).


\bibitem{Tonry:1981}
J. L. Tonry,
Astrophys. J. \textbf{251}, L1-L5 (1981).


\bibitem{Peterson:1975}
C. J. Peterson and I. R. King,
Aston. J. \textbf{80}, 427-436 (1975).

\bibitem{Illingworth:1976}
G. Illingworth,
Astrophys. J. \textbf{204}, 73-93 (1976).




\bibitem{Forbes:2020}
D. Forbes, A. Alabi, A.~J. Romanowsky, J.~P. Brodie, and N. Arimoto,
Mon. Not. Roy. Astron. Soc. \textbf{492}, 4874 (2020).

\bibitem{vanDokkum:2019}
P. van Dokkum, S. Danieli, R. Abraham, C. Conroy, and A.~J. Romanowsky,
Astrophys. J. Lett., \textbf{874}, L5 (2019)

\bibitem{Pina:2019}
Pi\~{n}a et al.,  Astrophys. J. Lett. \textbf{883}, L33 (2019)



\bibitem{Kaplinghat:2001me} 
  M.~Kaplinghat and M.~S.~Turner,
  Astrophys.\ J.\  \textbf{569}, L19 (2002).

\bibitem{Navarro:2016bfs} 
  J.~F.~Navarro et al.,
  Mon.\ Not.\ Roy.\ Astron.\ Soc.\  \textbf{471}, 1841 (2017),
  Astrophys.\ Space Sci.\ Proc.\  \textbf{56}, 103 (2019).

\bibitem{Glowacki_2019}
M. Glowacki, E. Elson, and R. Dav\'{e},
Mon. Not. Roy. Astron. Soc., staa2616 (2020)
doi:10.1093/mnras/staa2616.






\bibitem{Khoury:2014tka}
J. Khoury, Phys. Rev. D \textbf{91}, 024022 (2015).

\bibitem{Verlinde:2017}
E.~P.~Verlinde, SciPost Phys.\  {\bf 2}, no. 3, 016 (2017).


\bibitem{Edmonds:2013hba}
  D.~Edmonds, D.~Farrah, C.~M.~Ho, D.~Minic, Y.~J.~Ng and T.~Takeuchi,
  Astrophys.\ J.\  {\bf 793}, 41 (2014);
  Int.\ J.\ Mod.\ Phys.\ A {\bf 32}, no. 18, 1750108 (2017),
  D.~Edmonds, D.~Farrah, D.~Minic, Y.~J.~Ng and T.~Takeuchi,
  Int.\ J.\ Mod.\ Phys.\ D {\bf 27}, no. 02, 1830001 (2017),
  D.~Edmonds, D.~Minic and T.~Takeuchi, in preparation.


\bibitem{Ho:2010ca}
  C.~M.~Ho, D.~Minic and Y.~J.~Ng,
  Phys.\ Lett.\ B {\bf 693}, 567 (2010);
  Gen.\ Rel.\ Grav.\  {\bf 43}, 2567 (2011)
  [Int.\ J.\ Mod.\ Phys.\ D {\bf 20}, 2887 (2011)];
  Phys.\ Rev.\ D {\bf 85}, 104033 (2012).





\bibitem{Zhang:2010qk} 
  Y.-Y.~Zhang, H.~Andernach, C.~A.~Caretta, T.~H.~Reiprich, H.~Boehringer, E.~Puchwein, D.~Sijacki and M.~Girardi,
  Astron.\ Astrophys.\  {\bf 526}, A105 (2011).


\bibitem{HIFLUGCS}
T. H. Reiprich and H. B\"{o}hringer,
Astrophys. J. {\bf 567}, 716 (2002).


\bibitem{Belli:2014}
S. Belli, A. B. Newman, and R. S. Ellis,
Astrophys. J. {\bf 783}, 117 (2014).

\bibitem{Meylan:1993yd}
G. Meylan and C. Pryor, in
{\it Structure and Dynamics of Globular Clusters}, Astronomical
Society of the Pacific Conference Series,
Vol. 50, edited by S. Djorgovski and G. Meylan (San
Francisco, California, 1993) pp. 31-64.
C. Pryor and G. Meylan, ibid, 
pp. 357-371.



\bibitem{King:1968}
I. R. King, E. Hedemann Jr., S. M. Hodge, and
R. E. White, 
Astron. J. {\bf 73}, 456 (1968).


\bibitem{Sanders:1994}
R. H. Sanders, Astron. Astrophys. {\bf 284}, L31 (1994),
Astrophys. J. {\bf 512}, L23 (1999).


\bibitem{Sanders:2010aa}
R. H. Sanders, Mon. Not. Roy. Astron. Soc. {\bf 407}, 1128 (2010).







\bibitem{Duranzo2017}
R. Duranzo, X. Hernandez, B. Cervantes Sodi, and S. F. S\'{a}nchez,
Astrophys. J. {\bf 837}, 179 (2017),\\
X. Hernandez and A. J. Lara-D.I.,
Mon.\ Not.\ Roy.\ Astron.\ Soc. {\bf 491}, 272 (2020).


\bibitem{Bilek2019}
M. B\'{i}lek, S. Samurovi\'{c}, and F. Renaud,
Astron. Astrophys. {\bf 629}, L5 (2019).



\bibitem{Freidel:2013zga}
L.~Freidel, R.~G.~Leigh and D.~Minic,
  Phys.\ Lett.\ B {\bf 730}, 302 (2014);
  Int.\ J.\ Mod.\ Phys.\ D {\bf 23}, no. 12, 1442006 (2014);
  JHEP {\bf 1506}, 006 (2015);
  Int.\ J.\ Mod.\ Phys.\ D {\bf 24}, no. 12, 1544028 (2015);
  Phys.\ Rev.\ D {\bf 94}, no. 10, 104052 (2016);
  J.\ Phys.\ Conf.\ Ser.\  {\bf 804}, no. 1, 012032 (2017);
  JHEP {\bf 1709}, 060 (2017);
  Phys.\ Rev.\ D {\bf 96}, no. 6, 066003 (2017);
  Int.\ J.\ Mod.\ Phys.\ A {\bf 34}, no. 28, 1941004 (2019),\\
L.~Freidel, J.~Kowalski-Glikman, R.~G.~Leigh and D.~Minic,
  Phys.\ Rev.\ D {\bf 99}, no. 6, 066011 (2019).



\bibitem{rBHM8}
P.~Berglund, T.~Hubsch and D.~Minic,
  Phys.\ Lett.\ B {\bf 798}, 134950 (2019);
Int.\ J.\ Mod.\ Phys.\ D {\bf 28}, 1902003 (2019);
JHEP {\bf 1912}, 166 (2019)
  [JHEP {\bf 2019}, 166 (2020)].




\bibitem{Minic:2020oho} 
  D.~Minic,
  in \textit{the Proceedings of the 10th Mathematical Physics Meeting: School and Conference on Modern Mathematical Physics},
  9-14 September 2019, Belgrade, Serbia
  [arXiv:2003.00318 [hep-th]].



\bibitem{Barnes:2017}
D. J. Barnes et al., 
Mon. Not. Roy. Astron. Soc. {\bf 471}, 1088 (2017).







\end{thebibliography}
%

\end{document}